\documentclass[aps,twocolumn,floatfix,amsmath,showpacs]{revtex4}
\usepackage{graphicx}

\newcommand{\refcite}[1]{Ref.~\onlinecite{#1}}
\newcommand{\I}{\refcite{VanZonSchofield02a}}
\newcommand{\refsec}[1]{Sec.~\ref{#1}}
\newcommand{\Refsec}[1]{Sec.~\ref{#1}}
\newcommand{\average}[1]{\langle #1\rangle}
\newcommand{\inprod}[2]{\langle #1\,#2^*\rangle}
\newcommand{\eql}[1]{\label{eq:#1}}
\newcommand{\eq}[1]{Eq.~\ref{eq:#1}}
\newcommand{\fig}[1]{Fig.~\ref{#1}}
\newcommand{\fastterm}[1]{\mbox{ fast term in $#1$}}
\newcommand{\vek}{\mathbf k}
\newcommand{\veq}{\mathbf q}
\newcommand{\barM}{{\bar M}}
\newcommand{\calP}{\mathcal P}
\newcommand{\calH}{\mathcal H}
\newcommand{\calL}{{\cal L}}
\newcommand{\calfastL}{\calP^\perp{\cal L}}
\newcommand{\ds}{\displaystyle}

\begin{document}

\newcommand{\mytitle}{Mode-coupling theory for multiple-time
correlation functions of tagged particle densities and dynamical
filters designed for glassy systems}

\newcommand{\myauthor}{Ramses\ van Zon$^{\dagger *}$ and Jeremy\ Schofield$^*$}

\newcommand{\myaddress}{$^\dagger$The Rockefeller University, 1230
York Avenue, New York, NY 10021-6399, USA\\ 
$^*$Chemical Physics Theory Group, Department of Chemistry\\
University of Toronto,Toronto, Ontario, Canada M5S 3H6}

\newcommand{\myabstract}{The theoretical framework for higher-order
correlation functions involving multiple times and multiple points in
a classical, many-body system developed by Van Zon and Schofield
[Phys. Rev.~E~{\bf 65}, 011106 (2002)] is extended here to include
tagged particle densities.  Such densities have found an intriguing
application as proposed measures of dynamical heterogeneities in
structural glasses.  The theoretical formalism is based upon
projection operator techniques which are used to isolate the slow time
evolution of dynamical variables by expanding the slowly-evolving
component of arbitrary variables in an infinite basis composed of the
products of slow variables of the system.  The resulting formally
exact mode-coupling expressions for multiple-point and multiple-time
correlation functions are made tractable by applying the so-called
$N$-ordering method.  This theory is used to derive for moderate
densities the leading mode coupling expressions for indicators of
relaxation type and domain relaxation, which use dynamical filters that
lead to multiple-time correlations of a tagged particle density.  The
mode coupling expressions for higher order correlation functions are
also succesfully tested against simulations of a hard sphere fluid at
relatively low density.}

\newcommand{\mypacs}{61.20.Lc, 05.20.Jj, 05.40.-a, 61.20.Ja}

\newcommand{\mydate}{\today}

\title{\mytitle}
\author{\myauthor}
\affiliation{\myaddress}
\date{\mydate}
\pacs{\mypacs}
\begin{abstract}
\myabstract
\end{abstract}
\maketitle

\section{Introduction}

A complete understanding of the physical processes underlying the
transition between the high-temperature exponential relaxation of
density fluctuations of a fluid and the non-exponential relaxational
profiles observed at lower temperatures, especially near the glass
transition, remains elusive.  A number of interesting features have
been noted in dense, supercooled systems from computer simulation
studies\cite{Kobetal97,YamamotoOnuki98,YamamotoOnuki98b,DoliwaHeuer98,Kob99,YamamotoOnuki99,SjodinSjolander78}
as well as multi-dimensional NMR\cite{Boehmeretal96,Trachtetal98} and
video microscopy experiments\cite{Marcusetal99,Weeksetal00} that
appear to be related to this cross-over from simple exponential to
multi or stretched exponential relaxation: One typical feature of such
systems is the appearance of heterogeneously-distributed regions of
the fluid which differ dramatically in their mobility and local
density.  While fluid motions are relatively unrestricted in regions
of low density, structural rearrangements in regions of high local
density at a given time have been observed to occur through relatively
rapid, collective string-like motions\cite{Kobetal97}.  Furthermore,
regions which at one time were of relatively low density and in which
particle motions were primarily fluid-like can become locally-dense
and immobile.

It is well-known that a variety of different mechanisms are consistent
with non-exponential relaxation and that this relaxation is somehow
related to the heterogeneous nature of the dense fluid.  In one
possible scenario of non-exponential relaxation, fluctuations of local
density relax in the same intrinsically non-exponential way, where the
cooperative motion of particles depends strongly on the local
environment and is correlated over a long time period or ``history''.
Another possible mechanism is that each region of the fluid rearranges
in a less strongly-correlated fashion but with different rates, so
that the observed non-exponentiality is a consequence of the
superposition of different exponential relaxation processes.  Of
course neither scenario may apply over all time scales and the
mechanism may shift from one that is primarily homogeneous to one that
is primarily heterogeneous.  It is equally possible that one type of
relaxation may not even predominate over another.

In order to distinguish between these scenarios it is helpful to
construct quantitative measures which unambiguously signal the
presence of a specific mechanism.  Such constructions can be based on
filters\cite{Heuer97,HeuerOkun97,DoliwaHeuer98,DoliwaHeuer99} that
select out sub-ensembles of particles to have specific dynamical
properties over a sampling period.  A simple example of such a filter
is one that selects out individual particles that move either more or
less than a critical distance over a fixed period of time.  One can
then examine time correlations within these sub-ensembles to gain new
insight into detailed features of the dynamics.  When time-filters are
utilized in this fashion, time correlation functions of particles
contained in the sub-ensemble necessarily involve multiple-time
intervals.  Filters based on single-particle properties then can be
expressed as multiple-time correlation functions of tagged particle
densities.

Given this interesting application of multiple-time correlation
functions, the need for a theory that enables one to calculate such
quantities from first principles is clear.  In
\refcite{VanZonSchofield02a} such a theory was developed based
on mode-coupling theory for multiple-time correlation functions of
collective densities (i.e., particle number, momentum and energy
densities).  The theory was tested successfully on a hard sphere gas
at moderate densities in the hydrodynamic limit where the relevance of
mode-coupling theory is well-established.  In this article, the theory
is extended to include multiple-time correlation functions of
arbitrary type, encompassing correlation functions involving only
tagged densities, only collective densities as well as any combination
of the two types.

The paper is organized as follows: The mode-coupling formalism of
multiple-time correlation functions is introduced in
\refsec{Formalism} based on projection operator methods, and
equations describing the evolution of multiple-point densities of
arbitrary type are obtained.  These equations are manipulated to yield
expressions for correlation functions of multiple-time intervals that
are local in all time arguments.  In \refsec{Norder}, a
systematic method of determining what types of contributions are the
most significant for a particular multiple-time interval correlation
function of tagged particle densities is introduced and the
leading-order expressions for two and three-time interval correlations
of a tagged particle density are presented.  In \refsec{Applications},
specific applications of multiple-time correlation functions of tagged
particle densities are examined.  In particular, direct measures of
relaxation type and of the rate at which solid-like domains become
fluidized are analyzed and leading order mode-coupling expressions for
these measures are obtained.  \Refsec{Applications} concludes with
a numerical comparison of the leading order theoretical predictions
with direct simulation results for a low density, hard sphere system
in the hydrodynamic limit and \refsec{Summary} contains a summary.

\section{Mode-coupling formalism}
\label{Formalism}

\subsection{Slow variables}
\label{Slow variables}
The basic assumption of mode-coupling theory is that the long time
behavior of all time correlations functions can be completely
expressed in terms of the evolution of a set of slow modes of the
system. Although the theory doesn't specify the identity of the slow
modes, physical arguments can often serve as a guide to define a
finite set of variables that serve as a dynamical basis for the long
time evolution.  For example, since the particle number, the momentum
and the total energy of a fluid of structureless particles are
constants of motion, a minimal basis set for the slow evolution of the
system must include the long wavelength modes of densities of these
quantities.

Once the slow basis set has been determined, the slow component of an
arbitrary dynamical quantity can be extracted by finding the
projection of the variable onto the subspace spanned by the set of
slow variables.  Similarly, projecting the variable onto the
complement of this subspace should yield a fast quantity.

Projection operators are linear operators, and hence only the linear
dependence of a dynamical variable on the slow variables can be
projected out by such a procedure.  In general, however, one expects
the time dependence of most dynamical variables to be a {\it
non-linear} function of the slow variables.  This non-linearity can be
incorporated into the theory by constructing the slow subspace to
include the space spanned by all powers of the slow variables. In this
way an analytic dependence of the slowly-evolving component of a
dynamical variable on the slow variables can be described.  The basis
of the slow subspace is referred to as the multi-linear basis.

Consider an equilibrium system composed of $N$ particles that is
described by a translationally invariant Hamiltonian $\calH$. The time
evolution of any quantity (phase space function) $C$ is given by
\begin{equation}
	\dot C(t) = \calL C(t),
\eql{Liouville}
\end{equation}
with $\calL$ the Liouvillian operator, which is the Poisson bracket
with the Hamiltonian.

To describe the slow evolution of dynamical variables within the
projection operator formalism, the slow variables of the system are
taken together in a single vector $B$. As mentioned above, for a
structureless fluid, these are typically the number density, the
momentum density and the energy density:
\[
B(\mathbf r,t) = \left(\begin{array}{c}
		N(\mathbf r,t)\\
		\mathbf P(\mathbf r,t)\\
		E(\mathbf r,t)
	      \end{array}\right)
	    = \left(\begin{array}{c}
		\ds\sum_{n=1}^N\delta(\mathbf r-\mathbf r_n(t)) \\
		\ds\sum_{n=1}^N \mathbf p_n(t) \delta(\mathbf r-\mathbf r_n(t))\\
		\ds\sum_{n=1}^N e_n(t) \delta(\mathbf r-\mathbf r_n(t))
	      \end{array}\right).
\]
Here, $\mathbf r_n(t)$ is the position of particle $n$ at time $t$,
$\mathbf p_n(t)$ is its momentum and $e_n(t)$ its energy (kinetic and
potential).
It will be  convenient to work in Fourier space:
\begin{eqnarray}
B(\vek,t) &=& \sum_{n=1}^Ne^{i\mathbf k\cdot\mathbf r_n(t)} b_n(t),
\eql{Bset}
\end{eqnarray}
where $b_n(t)=(1,\mathbf p_n(t),e_n(t))$.

Since the goal of the mode-coupling theory outlined here is to
describe the time correlation functions of single particles, slow
tagged particle densities must be included, i.e.,
\begin{eqnarray*} 
	N_1(\mathbf r,t)&=&\delta(\mathbf r-\mathbf r_1(t)),
\end{eqnarray*} 
where particle $1$ is the tagged particle. In the Fourier representation, the 
tagged particle number density is simply
\begin{eqnarray} 
	N_{1}(\vek)= e^{i\mathbf k\cdot\mathbf r_1(t)}.
\eql{Ndef}
\end{eqnarray} 
The tagged particle density is taken together with $B(\vek)$ into a
larger vector $A(\vek)$. Henceforth, components of $A$ will be denoted
with a superscript $s$ when referring to {\em single particle}
densities (which have no summation over particles in their
definition) and with a superscript $c$ when involving {\em collective
densities} (which have a summation over all particles in their
definition).

We require that the slow subspace be spanned by all powers of the slow
variables. Since the product $N_{1}(\vek)N_{1}(\veq)=N_{1}(\vek+\veq)$
is just a linear variable, beyond the linear level only products of
$B$ are relevant.  Furthermore, to accommodate correlation
functions of certain fast variables one might be interested in
(e.g. $\mathbf P_{1}(\vek) = \mathbf p_1(t) e^{i\mathbf k\cdot{\mathbf
r}_1(t)}$), $A'(\vek)$ is defined as the vector composed of all linear
slow variables plus any fast variables of interest.

It will prove to be convenient to construct a basis set that is
orthogonal in the number of factors of the linear set $A$ and $B$,
which is guaranteed by the definition:
\begin{eqnarray*}
	Q_0 &\equiv& 1\\
	Q_1(\mathbf k) &\equiv& A'(\vek) - \average{A'(\vek)}  
	= \hat A'(\vek)\\
	Q_2(\mathbf k_1,\mathbf k_2)
	&\equiv& \hat A({\mathbf k_1})\hat B({\mathbf k_2} ) \\
	&& -\sum_{|\alpha|=0}^1
	\inprod{\hat A({\mathbf k_1}) \hat B({\mathbf k_2})}{Q_\alpha}
	*K_{\alpha\hat\alpha}^{-1}*Q_{\hat\alpha}\\ 
	&\vdots&
\end{eqnarray*}
Here the following notation has been used:
\begin{itemize}

\item $\langle \cdots \rangle$ denotes the (grand canonical) equilibrium
ensemble average, which is used to define the inner product.

\item A superscript ``$*$'' defines complex conjugation.

\item A Greek lower case letter denotes a set of pairs, each pair
containing a component index and a wave vector.

\item $|\alpha|$ denotes the number of pairs in the set
$\alpha$, the so-called {\em mode order}.

\item A hatted Greek letter has the same mode order as its unhatted
variant but is otherwise unrelated.

\item $Q_{\alpha}$ is the same as $Q_{|\alpha|}(\mathbf
k_1,\ldots,{\mathbf k}_{|\alpha|})$.  Also, in the rest of the paper,
the short hand notation of a number $n$ instead of a Greek letter will
often be used to indicate a set of mode order is n.

\item The ``$*$''product involves a summation over the pairs of wave
vectors and indices in $\alpha$, divided by all ways in which these
pairs can be permuted, so as to avoid over-counting.
Also, for this summation to always be well-defined, the wave vectors
will be summed up to a cut-off $k_c$, thus including $M=O(N)$ wave
vectors.

\item By definition,
$K_{\alpha\beta}\equiv\inprod{Q_\alpha}{Q_\beta}$, while $K^{-1}$ is
the inverse of this object with respect to the ``$*$'' product.

\end{itemize}
As the definition of $Q_n$ contains $K_{mm}$ for all $m<n$, which are
defined using the $Q_{m<n}$, the above definition of the multi-linear
basis is really recursive.  Note that as $K_{nm}=0$ if $n\neq m$, the
above set is orthogonal in mode order: $K_{nm}=\average{Q_n Q_m}=0$
unless $n=m$.

We note also that $\hat N_1(\vek=0)=(\exp[i\vek\cdot\mathbf
r_1]-\delta_{\vek0})\big|_{\vek=0}=0$. This implies that
$\average{\hat N_1(0,t)\hat N^*_1(0)}=0$ and, more importantly, that
the elements $Q_\alpha$ which have a $N_1$ component with zero wave
vector have to be omitted from the set to avoid an over-complete
basis.

\subsection{Single time interval correlations}
\label{stcorr}
Let all time correlation functions of the slow (linear) variables be
taken together into the matrix
\begin{equation}
	G_{11}(t) = \inprod{Q_1(t)}{Q_1} * K_{11}^{-1}
\end{equation}
Note that $G_{11}$ is wave vector dependent, but that this is not
explicitly denoted here.

In order to make contact with other versions of mode-coupling theory,
we first examine the consequences of taking only the linear basis
$Q_1$ into account. The time evolution of $Q_1$ is given by
\eq{Liouville}. In differential form, this means
\begin{equation}
	\dot Q_1(t) = \calL\,Q_1(t) = e^{\calL t} \calL\,Q_1.
\eql{Adiff}
\end{equation}
Defining the linear projection operator
\begin{eqnarray*}
	\calP_1 C &\equiv & \inprod{C}{Q_1}*K_{11}^{-1}* Q_1 
\end{eqnarray*}
and its complement $\calP_1^\perp \equiv 1 - \calP_1$, and using a
well known operator identity, \eq{Adiff} can be cast in the form of a
generalized Langevin equation
\begin{equation}
	\dot Q_1(t) = M^{E}_{11} * Q_1(t) + \int_0^t
	\tilde{M}_{11}(t-\tau) *
	Q_1(\tau) \,d\tau + \tilde\varphi_1(t),
\eql{memorydiff}
\end{equation}
in which a {\em fluctuating force} $\tilde\varphi_1(t) \equiv \ds
e^{\calP_1^\perp\calL t}\calP_1^\perp\calL\,Q_1$, a {\em static
vertex} $M^{E}_{11} \equiv \inprod{ \{\calL\,Q_1\} }{Q_1} *
K_{11}^{-1}$ and a {\em memory kernel} $\tilde M_{11}(t) \equiv
-\inprod{\tilde\varphi_1(t)}{\tilde\varphi_1} *K_{11}^{-1}$ appear.
By taking the inner product with $Q_1$, \eq{memorydiff} yields
\begin{equation}
	\dot G_{11}(t) = M^{E}_{11} * G_{11}(t) 
	+ \int_0^t \tilde{M}_{11}(t-\tau) 
	* G_{11}(\tau) \,d\tau.
\eql{memorycorr}
\end{equation}

It should be noted that because of translational symmetry, the sum of
wave vectors in an average has to add up to zero. Thus $G_{11}$
involves only one wave vector, instead of two. Likewise, the above
equation is an equation involving just one wave vector.  In this
sense, there is no mode-coupling in the above equation, although, of
course, the different components of $Q_1$ are coupled.

The memory kernel $\tilde{M}_{11}(t)$ involves the time correlation of
the fluctuating force $\tilde\varphi_1$, for which the formalism as
presented thus far provides no method of evaluating.  But provided
$\calP_1^\perp$ projects out all the slow behavior of $\varphi(t)$,
$\inprod{\tilde\varphi_1(t)}{\tilde\varphi_1}$, which contains
dynamics orthogonal to $\hat A$ only, should be a quickly-decaying
function that is well-approximated at long times $t \gg t_m$ by
$\tilde M_{11}(t) \approx 2D \delta(t)$, for some microscopic time
$t_m$ on the order of the particle-particle collision time.  Under
such circumstances, \eq{memorycorr} is local in time, i.e., $\dot
G_{11}(t) = (M^{E}_{11}+D) * G_{11}(t)$.  Unfortunately, in most
cases, the correlation function of the dissipative force
$\tilde\varphi_1(t)$ is not a quickly-decaying function but instead
has long time
tails\cite{AlderWainwright68,AlderWainwright69,MichaelsOppenheim75}
due to the fact that the linear projection operator $\calP_1^\perp$
does not remove the non-linear dependence of $\tilde{\varphi}(t)$ on
the set of slow variables $\hat B$.  Hence, one is forced to use the
full multi-linear basis if equations of motion are to be local in
time.

In the following, $Q$ will denote the vector composed of all
$Q_\alpha$. As $Q$ is still a phase space function, its time evolution
is governed by
\[
	\dot Q(t) = \calL\,Q(t).
\]
Using the multi-linear projection operator
\begin{eqnarray*}
	\calP C &\equiv& \inprod{C}{Q} *
	K^{-1} * Q,\\
\end{eqnarray*}
(where the ``$*$'' now also denotes a sum over mode-orders) and its
complement $\calP^\perp\equiv 1 - \calP$, an equation analogous to
\eq{memorycorr} is found:
\begin{eqnarray}
	\dot G(t) &=& M^{E} * G(t) 
	+ \int_0^t M^D(t-\tau) 
	* G(\tau) \,d\tau,
\eql{Qmemorycorr}
\end{eqnarray}
where $G(t)$ is the full propagator defined by
\begin{equation}
	G(t) \equiv \inprod{Q(t)}{Q} * K^{-1},
\eql{Gsingledef}
\end{equation}
the fluctuating force is defined by
\begin{equation}
	\varphi(t) \equiv e^{\calP^\perp\calL t}\calP^\perp\!\calL\,Q, 
\eql{fluctuatingForce}
\end{equation}
and the vertices  $M^E$ and $M^D(t)$ are given by
\begin{eqnarray*}
	M^{E} &=& \inprod{\{\calL\,Q\}}{ Q}*K^{-1},\\
 	M^{D}(t) &=& -\inprod{\varphi(t)}{\varphi}*K^{-1}.
\end{eqnarray*}
Note that the original goal of evaluating $G_{11}(t)$ now becomes to
calculate the $1-1$ (or linear-linear) sub-block of
$G(t)$ while  $G_{0\alpha}(t)=G_{\alpha0}(t)=\delta_{\alpha0}$ is trivial.

Now that all powers of $\hat B$ are projected out of the dynamics, one
expects $\varphi$ to be truly fast, and its correlation function to be
approximately a delta function. Defining the dissipative vertex as
\begin{eqnarray*}
	M^D &\equiv& \int_0^{\infty}\!\! M^D(t) \,dt,
\end{eqnarray*}
and the full vertex to be
\begin{eqnarray}
	M \equiv M^E + M^D 
\eql{Vertices}
\end{eqnarray}
we can write 
\begin{equation}
	\dot G(t) = M * G(t),
\eql{Qinstantt}
\end{equation}
as an approximation to \eq{Qmemorycorr}.

A physical note is in order here: it is assumed that there is a
separation of time scales, such that there are fast correlations which
decay on the short microscopic scale $t_m$, while the interesting long
time behavior occurs on slow, `hydrodynamic' scales $t_h$, and we
assume $t_m\ll t_h$. \eq{Qinstantt} describes just the slow part, so
it is valid only after a time (much) larger than $t_m$ with
corrections of order $O(t_m/t_h)$.

It is useful to perform a Laplace transform:
\[
	G(z) = \int_0^\infty e^{-zt} G(t) \,dt.
\]
With the initial condition $G_{\alpha\beta}(t=0)=1_{\alpha\beta}$
(the unit matrix in infinite dimensions),
\eq{Qinstantt} can be solved formally in Laplace space,
\[
	G(z) = [ z - M ]^{-1},
\]
where the inverse is to be taken with respect to the ``$*$'' product.
By splitting up the matrix $M_{\alpha\beta}$ into its part diagonal in
wave vectors $M^d$ (i.e., the wave vectors in $\alpha$ and $\beta$ are
pair-wise equal) and an off-diagonal remainder, $M^o=M-M^d$, one can
write
\begin{equation}
	G(z) = [ z - M^d - M^o ]^{-1} = G^b(z) * [1 - M^o G^b(z)]^{-1},
\eql{inverse}
\end{equation}
where the bare, mode-order diagonal propagator $G^b$ is defined as
\[
	G^b(z) \equiv [z-M^d]^{-1}.
\]
Because the wave vectors in $\alpha$ and $\beta$ have to pair up in
$M^d_{\alpha\beta}$, the numbers of wave vectors $|\alpha|$ and
$|\beta|$ need to be equal as well, i.e., $M^d$ and $G^d$ are also
diagonal in mode-order.  $G^b(z)$ at a particular mode-order $n$ is
denoted as $G^b_{n}(z)$.

The inverse in \eq{inverse} can be expanded to yield
\begin{eqnarray}
	G(z) &=& G^b(z) + G^b(z) *M^o *G^b(z) \nonumber\\&&
+ \, G^b(z) *M^o *G^b(z)* M^o*G^b(z)
+\ldots \eql{Gexpand}
\end{eqnarray}
The $1-1$ element of $G(z)$ can be written in terms of a {\em
self-energy} that is defined as
\begin{eqnarray}
\Sigma(z) &\equiv&  \sum_{n=2}^{\infty} M^o_{1n}* G_{n}^b(z) *M^o_{n1}
\nonumber\\
&+&
  \sum_{n=2}^\infty \sum_{m=2}^\infty M^o_{1n}*  G_{n}^b(z) *M^o_{nm}
*G_{m}^b(z) *M^o_{m1}
\eql{selfenergy}\\
&+&\ldots\nonumber
\end{eqnarray}
Note that the summations start at mode-order $2$. 
After re-summing \eq{selfenergy}, $G_{11}(z)$ is related to $\Sigma(z)$ by
\begin{eqnarray}
	G_{11}(z) &=& [z - M_{11} - \Sigma(z)]^{-1}.
\eql{G11z}
\end{eqnarray}
This result can be compared to the solution in Laplace space
of \eq{memorycorr}, $G_{11}(t) = [z-M^E_{11}-\tilde M(z)]^{-1}$.
Thus, the self energy is related to the memory kernel by
\begin{equation}
	\tilde M(z) = M^D_{11} + \Sigma(z).
\eql{MemoryKernel}
\end{equation}
Since $M^D_{11}$ is short lived, the long time tails of the memory
function are due to mode-coupling effects in $\Sigma(z)$.

It was shown by Schofield and Oppenheim\cite{SchofieldOppenheim92}
that in the thermodynamic limit, the series for the self energy for
$\Sigma$ can be re-summed, with the result that all bare propagators
in \eq{selfenergy} are replaced by propagators that are completely
diagonal in wave vector, but with the restriction of the sum over
intermediate wave vector sets that none of them should be equal. In
the same paper it was shown that that these full diagonal multi-linear
propagators factor in the thermodynamic limit into products of full
linear-linear propagators. As a result, \eq{selfenergy} and \eq{G11z}
combine to a self-consistent equation for $G_{11}(z)$.  It should be
noted that these results rely on the $N$-ordering technique, which
will be discussed in detail in section \ref{Norder}.

\subsection{Multiple-point correlations}
\label{mpcorr}
In \I, the functions $G_{\alpha\beta}(t)$ with either $|\alpha|$ or
$|\beta|$ bigger than one were called multiple-point correlation
functions because they can be seen as Fourier transforms of
correlation function of densities involving more than one relative
position.  Note that if $|\alpha|=|\beta|$ {\em and} the wave vectors
in $\alpha$ and $\beta$ are fully (pairwise) matched,
$G_{\alpha\beta}$ is the full propagator at mode-order $|\alpha|$, and
this propagator can be written as a product of full-linear propagators
to an excellent approximation. However, even when the wave vectors are
not fully matched, $G_{\alpha\beta}$ is still an interesting but
nontrivial quantity.

The expression for the multiple-point correlation functions follows
from the general form in \eq{Gexpand}. The form of the equation is
identical to the case of just collective densities that was treated in
\I\ (section II D). By performing the same re-summations, which rely
on the $N$-ordering method to be discussed shortly, one obtains the
result of Eq.~(26) of that paper:
\begin{eqnarray}
        G_{\alpha\beta} &=&
        G_{\alpha\alpha'}\delta_{\alpha' \beta}
        + G_{\alpha\alpha'}*M^o_{\alpha'\beta'}*
        G_{\beta'\beta}
\nonumber\\&&
        + \, G_{\alpha\alpha'}*M^o_{\alpha'\delta}*
          G_{\delta\delta'}*
          M^o_{\delta'\beta'}*
          G_{\beta'\beta}
\nonumber\\
&&
+ \ldots.
\eql{physicalexpansion}
\end{eqnarray}
Here, primed Greek indices have the same wave vectors as their
unprimed variants, but not necessarily the same component indices,
i.e., they are fully diagonal in wave vector.  Thus,
$G_{\alpha\alpha'}\delta_{\alpha' \beta}$ denotes the full diagonal in
wave vector of the propagator at mode-order $|\alpha|$.  Furthermore,
there is a restriction on the summation that none of the intermediate
wave vector sets are equal, which in contrast to the notation in \I\
is not denoted explicitly here.

Eq.~(\ref{eq:physicalexpansion}) expresses the multiple-point correlation
function in terms of the full propagators, which can be expressed in
terms of the full linear-linear propagators. By using $N$-ordering
(see section \ref{Norder}), one can show that contributions in
\eq{physicalexpansion} involving $G_n(z)$ with $n<|\alpha|$ and
$n<|\beta|$ are negligible in the thermodynamic limit.

\subsection{Multiple-time correlations}
The results above consider only correlation functions that contain a
single time interval. However, the case of correlation functions of
several time intervals, which in general will be denoted (following
\I) by
$G^{(n)}_{\alpha_n,\alpha_{n-1},\ldots,\alpha_0}(t_n,t_{n-1},\ldots,
t_{1})$ or
\[
 G^{(n)}_{\{\alpha_i\}}(\{t_i\}) \equiv
\average{
Q_{\hat\alpha_0}^* Q_{\alpha_n}(t_1+\cdots+t_n) \ldots
Q_{\alpha_1}(t_1)} * K^{-1}_{\hat{\alpha}_0\alpha_0}
,
\]
is of considerable interest. 

Even very straightforward approaches to multiple-time correlation
functions lead to expressions involving multiple-time correlations of
the fluctuating force. Such correlation functions are generally
non-trivial as they are not always fast
decaying\cite{SchrammOppenheim86}.  The essential ingredient in
deriving mode-coupling expressions for multiple-time correlation
functions, as remarked in \refcite{VanZonSchofield02a}, is
that `` {\em in a correlation function involving fluctuating forces,}
{\em the function decays quickly in a pair of time arguments,} {\em
provided these are well-separated in time}.'' Here, ``well-separated''
means having a time separation much larger than the microscopic time
scale $t_m$ on which $\inprod{\varphi(t)}{\varphi}$ decays. For
correlation functions of collective densities, this argument led to
the conclusion that a correlation function involving several time
arguments $t_i$ whose time evolution arises through the fast
evolution operators $\exp[{\calP^\perp\calL t_i}]$ can be considered
fast-decaying in each of the times $t_i$ provided all times are
positive, larger than $t_m$, and the evolution operators are applied
in succession. When these properties hold, all effective times (such
as $t_1$, $t_1+t_2$, $t_1+t_2+t_3$, etc.) are well-separated and the
quoted criterion applies.

On a formal level, the arguments invoked in \I\ to derive expressions
for multiple-time correlation functions of collective densities apply
equally well when tagged particle densities are included in the slow
basis set. Thus, by using projection operator techniques, the
following relation can be established:
\begin{eqnarray*}
G^{(n)}_{\{\alpha_i\}}(\{t_i\})
&=&
 \int_0^{t_{n-1}}\!\!\!\int_0^{t_n}\!\!\!\,
        G_{\alpha_n\beta}(t_n-\tau)
        M_{\beta\alpha_{n-1}\delta}(\tau,\tau_1)
\nonumber
\\
&&
        \times
        G^{(n-1)}_{\delta,\alpha_{n-2}\ldots}(t_{n-1}-\tau_1,t_{n-2},\ldots) \;
         d\tau d\tau_1
\nonumber\\&&
 + \fastterm{t_{n-1}},
\end{eqnarray*}
where
\begin{subequations}
\eql{bothparts}
\begin{align}
        M_{\beta\alpha\delta}(\tau,\tau_1) =& \,
        4\average{Q_\beta Q_\alpha
        Q_{\hat\delta}^{*}} K^{-1}_{\hat{\delta}\delta} \delta(\tau)\delta(\tau_1)
\eql{part1} \\
&
        -\average{
        \dot{Q}^*_{\hat{\delta}}
        e^{\calfastL\tau_1}\calP_\perp
        \varphi_\beta(\tau) Q_\alpha }
        K^{-1}_{\hat{\delta}\delta}
\eql{part2}.
\end{align}
\end{subequations}
In the limit of long times where $\tau_1 \gg t_m$, the fast term can be
neglected and the integrand can be replaced by a delta function
according to the above criterion, yielding an equation that is local
in $t_n$,
\begin{equation}
        G^{(n)}_{\alpha_n,\ldots}
        (t_n,\ldots)
=
  G_{\alpha_n\beta}(t_n)
        \barM_{\beta \alpha_{n-1}\delta}
        G^{(n-1)}_{\delta,\alpha_{n-2},\ldots}(t_{n-1},\ldots)
\eql{instantmultiQ},
\end{equation}
where $\barM_{\delta \alpha\theta}= \int_0^\infty d\tau_1
\int_0^\infty d\tau M_{\delta \alpha\theta}(\tau,\tau_1)$.  Different
from the single time correlations of sections \ref{stcorr} and
\ref{mpcorr}, where including $Q_0$ was of little practical
consequence, for the multiple-time correlations here it may happen
that $|\beta|$ or $|\delta|$ is zero. For these special cases,
$\barM_{\beta\alpha\delta}$ are given by
$\barM_{\beta\alpha0}=K_{\beta\alpha^*}$ and
$\barM_{0\alpha\delta}=1_{\alpha\delta}$.

The recursion relation in \eq{instantmultiQ} can be applied as many
times as necessary to yield a relation between $G^{(n)}$ and
$G^{(1)}$. For instance, for $n=2$ and $n=3$:
\begin{eqnarray}
 G^{(2)}_{\alpha \gamma\beta}(t_2,t_1)
&=&G_{\alpha\delta}(t_2)
\barM_{\delta \gamma\theta}  G_{\theta\beta}(t_1)
\eql{instantQQQ}\\
G^{(3)}_{\alpha\beta\gamma\delta}(t_3,t_2,t_1) &=&
        G_{\alpha\zeta}(t_3)\barM_{\zeta\beta\theta}
        G_{\theta\eta}(t_2)\barM_{\eta\gamma\lambda}
        G_{\lambda\delta}(t_1).
\nonumber
\end{eqnarray}

\section{$N$-ordering scheme for tagged and mixed correlations}
\label{Norder}

In order to make the infinite series such as in \eq{selfenergy} or in
\eq{instantmultiQ} tractable, one must develop a systematic scheme for
analyzing the relative importance of terms appearing in the series.
In such an approach, the series must be analyzed so that simple but
accurate approximations for the entire series can be formulated in a
(preferably) controlled fashion.  The $N$-ordering method, developed
by Oppenheim and co-workers\cite{MachtaOppenheim82,Schofieldetal92} as
an extension of Van Kampen's system size expansion\cite{VanKampen},
allows such an approach for correlation functions on the hydrodynamic
length scale in systems of moderate density removed from any critical
point.  In the $N$-ordering approach, one assigns a factor of $N$ (the
number or average number of particles) to any cumulant of multi-linear
densities based on the assumption that the each cumulant containing
$n$ linear densities scales with the system size as
$N(\xi/a)^{3(n-1)}$, where $\xi$ is the correlation length of the
system and $a$ is the average distance between particles.  As shown in
\refcite{SchofieldOppenheim92}, the $N$-order of an
arbitrary correlation function of basis functions $K_{\hat{\alpha}
\alpha} = \left\langle Q_{\hat\alpha} Q_{\alpha} \right\rangle$
depends on the nature of the densities and the number of matched wave
vectors in the sets $\alpha$ and $\hat{\alpha}$.  The subscripts like
\begin{equation}
\alpha = \{ A_1(k-q_1 - \cdots - q_{|\alpha|-1}), B_1(q_1), \dots , B_{|\alpha| - 1}(q_{|\alpha}| -1) \}
\nonumber
\end{equation}
denote sets of wave vector-dependent densities, where each $B_i$ is a
component of the set of collective slow variable defined in
\eq{Bset} and $A_i$ is a component of the extended linear basis
set $A$.  If the component $A_1$ in the set $\alpha$ is a single
particle density, we denote the corresponding multi-linear density by
a superscript $s$, $Q_{\alpha}^s$; if this component is a collective
density, then the dynamical variable is represented by $Q_\alpha^c$.
In \refcite{SchofieldOppenheim92}, it was demonstrated that
the leading $N$-order terms arise from matching as many wave vectors
as possible in the sets $\alpha$ and $\hat{\alpha}$, yielding the
following estimates for $K_{\hat{\alpha} \alpha}$,
\begin{eqnarray}
\begin{matrix}
K_{\hat{\alpha} \alpha}^{cc} \sim N^{|\alpha|} \, & \, K_{\hat{\alpha} \alpha}^{cs} \sim N^{|\alpha|-1} \\
K_{\hat{\alpha} \alpha}^{sc} \sim N^{|\alpha|-1} \, & \, K_{\hat{\alpha} \alpha}^{ss} \sim N^{|\alpha|-1}  
\end{matrix}
\eql{NorderK} 
\end{eqnarray}
where, for example, the superscript ``$ss$'' denotes that both sets
$\hat{\alpha}$ and $\alpha$ contain a single particle density.  More
importantly, it was also demonstrated that the $N$-order of
$K_{\hat{\alpha} \alpha}^{ss} \sim N^{|\alpha|-1}$ only when the
single particle densities are in the same matched set, and the
$N$-order drops by a factor of $N$ when the wave vectors of the single
particle densities do not match.

For the inverse of $K$, the $N$ ordering in \eq{NorderK} implies
\begin{eqnarray}
\begin{matrix}
(K^{-1})_{\hat{\alpha} \alpha}^{cc} \sim N^{-|\alpha|} \, 
& \, (K^{-1})_{\hat{\alpha} \alpha}^{cs} \sim N^{-|\alpha|} \\
(K^{-1})_{\hat{\alpha} \alpha}^{sc} \sim N^{-|\alpha|} \, 
& \, (K^{-1})_{\hat{\alpha} \alpha}^{ss} \sim 
N^{-|\alpha|+1}.
\eql{NorderKinverse}
\end{matrix}
\end{eqnarray}

When considering the $N$ order of an expression that contains the
``$*$'' product, one needs to separately consider its order in $M$
(the number of wave vectors summed over, see the seventh point at the
end of section \ref{Slow variables}). That is, when the ``$*$''
product is between multi-indices of order $n$, one has in principle
$n-1$ sums over wave vectors and thus an effective factor of
$M^{n-1}$ (one loses one sum over a wave vector because by
translation symmetry all wave vectors have to add up to zero inside an
average). Although $M$ is of order $N$, these orders of $M$ are
counted separately from the $N$ ordering, for the following good
reason: For fluids of moderate density only a small fraction of the
wave vectors in the sums really contributes significantly.
Therefore, rather than taking $M$ as counting the precise number of
wave vectors summed over, it makes more sense to take $M$ as the
number of wave vectors that contribute substantially to the sum.  This
results in a small value of $M$ [which is nonetheless still
$O(N)$]. The $O(1)$ parameter $M/N$ is called the mode-coupling
parameter. As stated, it is typically small (e.g.\ $10^{-5}$) for
fluids of moderate density and away from any critical point.

To illustrate this, consider for instance $K^{cc}_{2\hat
2}*K^{cc}_{\hat 2{\hat 2}'}$ which is seen to have an $N$-ordering of
$O(N^{4})$. There is a possible summation over $M$ wave vectors in
$K^{cc}_{2\hat 2}*K^{cc}_{\hat 2{\hat 2}'}$, but in fact the leading
order estimate of $O(N^{4})$ requires a matching of wave vectors which
gets rid of the factor $M$. As a result, the leading term of
$K^{cc}_{2\hat 2}*K^{cc}_{\hat 2{\hat 2}'}$ is just $O(N^{4})$, with
(possible) correction terms of order $O(MN^{3})$, which compared to
the leading term are of relative order $M/N$, i.e., of the mode
coupling parameter.

In general, for any expression one can determine the values of $m$ and
$n$ such that it is of order $O(M^mN^n)$=$O((M/N)^m N^{n-m})$. If
$m>n$, such an expression vanishes in the thermodynamic limit
$N\to\infty$. Below, when adding expressions of different $M$ and $N$
order, expressions which vanish (relatively) in the thermodynamic
limit will be omitted. The remaining terms with the smallest power of
$M/N$ (typical $(M/N)^0$) will be referred to as the {\em leading
$N$-order terms} and the terms of one power higher in $M/N$ will be
called the {\em leading correction terms} or the {\em first
mode-coupling corrections}.

\subsection{$N$-ordering of single time interval correlation functions}
Using these principles, it was shown\cite{SchofieldOppenheim92} that
the $N$-order of the (multi-linear) vertices in
\eq{Vertices} (in their maximally matched form) are:
\begin{eqnarray}
\left.\begin{array}{r}
   M^{ss}_{\delta\beta}\\
   M^{cs}_{\delta\beta}\\
   M^{cc}_{\delta\beta}
\end{array}\right\}
 &=& \left\{
	\begin{array}{lcl}
		O(N^{-(\beta-\delta)}) && \mbox{ if } \delta<\beta
\\
	O(N^{0}) && \mbox{ if } \delta\geq\beta
	\end{array}
	\right.
\nonumber\\
   M^{sc}_{\delta\beta} &=& N^{-1}O(M^{ss}_{\delta\beta}) \nonumber.
\end{eqnarray}
Following these lines of analysis, the matrix $M$ in
\eq{Vertices} and the normalized single-time interval
correlation functions in \eq{Gsingledef} are\cite{SchofieldOppenheim92}, in terms of
$N_{\alpha\beta}=\inprod{\{\calL\,Q_\alpha\}}{ Q_\beta} -\int_0^\infty
dt\,\inprod{\varphi_\alpha(t)}{\varphi_\beta}$
\begin{eqnarray}
M_{\alpha \beta}^{cc} &=& N_{\alpha \hat{\beta}}^{cc} * K_{\hat{\beta} \beta}^{cc-1} [1 + O(N^{-1})] \nonumber \\
M_{\alpha \beta}^{ss} &=& N_{\alpha \hat{\beta}}^{ss} * K_{\hat{\beta} \beta}^{ss-1} [1 + O(N^{-1})] \nonumber \\
G^{cc}_{\alpha \beta}(z) &=& \left[ z - M^{cc}(z) \right]^{-1}_{\alpha \beta} \nonumber \\
G^{ss}_{\alpha \beta}(z) &=& \left[ z - M^{ss}(z) \right]^{-1}_{\alpha
  \beta},
\eql{Gss}
\end{eqnarray}
from which we see that single particle modes do not contribute to the
dynamics of single time interval correlation functions of the
collective modes.  Similarly, since the
superscripts $ss$ in $M^{ss}$ in \eq{Gss} only mean that one of the
components in the each of the indices of $M^{ss}$ has a single
particle character while the others are collective, the collective
modes do influence the dynamics of correlation functions of tagged
(single) particle densities through mode-coupling corrections to
transport coefficients.  For instance, the generalized self-diffusion
coefficient $\tilde{D}(\vek,t)$ is renormalized through the terms in
\eq{selfenergy} with the bare propagator replaced by the full one (see
the discussion below \eq{MemoryKernel}), i.e.,
\begin{align}
\tilde{D}&(\vek,t) = D^{B}(\vek,t) + \Sigma^{ss}(\vek,t) \nonumber \\
\Sigma^{ss}&(\vek,t) = 
\sum_{i,j,\veq}
M_{12}^{ss}(N_1(\vek);N_1(\vek-\veq),B_i(\veq)) \, F_s(\vek-\veq;t) \nonumber \\
& \times G_{11}^{cc}(B_i(\veq);B_j(\veq);t) \,
M_{21}^{ss} (N_{1}(\vek-\veq),B_j(\veq);N_1(\vek)) 
\nonumber\\&
+ \cdots , \nonumber
\end{align}
where $D^{B}=M_{11}^{ss}$ 
is the ``bare'' diffusion coefficient 
with weak $\vek$ and
$t$ dependence and $F_s(k,t)=\langle \hat{N}_{1}(\vek,t)
\hat{N}_{1}^{*}(\vek) \rangle$ is the self-part of the dynamic
structure factor. Careful analysis of the mode-coupling contributions
to tagged particle transport coefficients is essential in the
derivation of observed relations between quantities such as the
self-diffusion coefficient of a Brownian particle and the viscosity of
the fluid\cite{KeyesOppenheim73,Keyes77,SchofieldOppenheim92}.

\subsection{$N$-ordering of higher-order vertices and correlation functions}
The analysis of the $N$-ordering of higher-order vertices involving
mixed tagged and collective particle densities follows by induction as
outlined in \refcite{VanZonSchofield02a}.  Using the
$N$-ordering results for the single time interval correlation
functions and the relation between the single-time interval
correlation functions and the multiple-time correlation functions in
\eq{instantQQQ}, one obtains the following $N$-order in the maximally
matched form of the higher order vertices in which the central index
is a linear tagged density:
\begin{eqnarray}
   \bar M^{csc}_{\gamma1\delta}
 &=& \left\{
	\begin{array}{lcl}
		O(N^{-(\delta-\gamma)}) && \mbox{ if } \gamma<\delta\\
		O(N^{-1}) && \mbox{ if } \gamma=\delta\\
		O(N^{0}) && \mbox{ if } \gamma>\delta
	\end{array}
	\right.
\nonumber\\
   \bar M^{ssc}_{\gamma1\delta}
 &=& \left\{
	\begin{array}{lcl}
		O(N^{-1-(\delta-\gamma)}) && \mbox{ if } \gamma<\delta\\
		O(N^{-1}) && \mbox{ if } \gamma=\delta\\
		O(N^{0}) && \mbox{ if } \gamma>\delta
	\end{array}
	\right.
\nonumber\\
   \bar M^{css}_{\gamma1\delta}
 &=& \left\{
	\begin{array}{lcl}
		O(N^{1-(\delta-\gamma)}) && \mbox{ if } \gamma<\delta\\
		O(N^{0}) && \mbox{ if } \gamma\geq \delta
	\end{array}
	\right.
\nonumber\\
   \bar M^{sss}_{\gamma1\delta} &=& O(M^{ss}_{\gamma\delta})\nonumber.
\eql{higherNorder}
\end{eqnarray}

Furthermore, the higher order propagators $G_{\gamma 1 \delta}$ obey
the same $N$-ordering rules as the higher-order vertices, namely
\begin{eqnarray}
	G_{\gamma1\delta} &=& O(\bar M_{\gamma1\delta}).
\end{eqnarray}

Using these results, we see that the two-time interval tagged particle
correlation function (with time intervals $t_1$ and $t_2$) reduces to
a particularly simple form to leading $N$-order:
\begin{eqnarray}
G^{(2)}_{111}(t_2,t_1) &=& \inprod{Q_{1}^{s}(\vek-\veq,t_1+t_2)Q_{1}^{s}(\veq,t_1)}{Q^{s}_{1}(\vek)} \nonumber \\
&=& G^{ss}_{11}(t_2)\bar{M}_{111}^{sss} G^{ss}_{11}(t_1) + O(N^{-1}),
\eql{leading}
\end{eqnarray}
with leading order correction terms
\begin{align}
G^{ss}_{12}(t_2)\bar{M}_{212}^{sss} G^{ss}_{21}(t_1)
+
 G_{12}^{ss}(t_2)\bar{M}_{211}^{sss}&G_{11}^{ss}(t_1)
\nonumber\\
 + G_{11}^{ss}(t_2)\bar{M}_{112}^{sss}G_{21}^{ss}(t_1)
\eql{correction}
\end{align}
where $G^{ss}_{12}(t)$ and $G^{ss}_{21}(t)$ are higher-order, single
time interval correlation functions.  Note that in \eq{leading} and
\eq{correction}, the matrix indices, as more fully indicated in
\eq{instantQQQ}, have been suppressed for notational simplicity.
Given the definition of $M_{112}$ in \eq{bothparts}, one sees that only
the second, dissipative term in \eq{part2} contributes, which is
$O(k_0^2)$. Thus the first term in \eq{correction} is, in orders of
$k_0$, the leading correction term, while the others are $O(k_0^2)$.

For the three-time interval correlation function, the leading
$N$-order contribution is
\begin{align}
G^{(3)}_{1111}&(t_3,t_2,t_1) = 
G^{ss}_{11}(t_3)\bar{M}_{110}^{ss}
G_{00}(t_2)\bar{M}_{011}^{\:\,ss} G^{ss}_{11}(t_1)
\nonumber\\
&+
G^{ss}_{11}(t_3)\bar{M}_{111}^{sss}
G^{ss}_{11}(t_2)\bar{M}_{111}^{sss} G^{ss}_{11}(t_1),
\eql{G3leading}
\end{align}
with the following three terms contributing at the order of the first
mode-coupling corrections:
\begin{align}
& G^{ss}_{12}(t_3)\bar{M}_{212}^{sss}
  G^{ss}_{21}(t_2)\bar{M}_{111}^{sss} G^{ss}_{11}(t_1)
\nonumber\\
&
+
 G^{ss}_{11}(t_3)\bar{M}_{111}^{sss}
  G^{ss}_{12}(t_2)\bar{M}_{212}^{sss} G^{ss}_{21}(t_1)
\nonumber\\
&+
 G^{ss}_{12}(t_3)\bar{M}_{212}^{sss}
  G^{ss}_{22}(t_2)\bar{M}_{212}^{sss} G^{ss}_{21}(t_1)
 + O(k_0^2).
\eql{G3mc}
\end{align}
The $O(k_0^2)$ correction here in fact consists of seven more
correction terms, all variants of the form
$G_{12}(t_3)\barM_{211}G_{11}(t_2)\barM_{111}G_{11}(t_1)$ and all
$O(k_0^2)$.

\section{Applications of higher order correlation functions}
\label{Applications}

It is often difficult using single time-interval correlation functions
to discern what features of the underlying collective dynamics of
slowly relaxing systems leads to qualitative features in the
relaxation profile in glassy and frustrated systems.  Typically,
single time-interval correlation functions in frustrated systems
display non-exponential time decay, often exhibiting a two-step
relaxation processes associated with caging effects and cooperative
flow through heterogeneous
dynamics\cite{YamamotoOnuki98,YamamotoOnuki98b,Trachtetal98} on long
time scales.  Given the heterogeneous nature of the dynamics in such
systems, it is natural to ask what {\it type} of relaxation processes
lead to this non-exponential time signature.

In a series of
articles\cite{Heuer97,HeuerOkun97,DoliwaHeuer98,DoliwaHeuer99}, Heuer
and co-workers have examined the information content of higher-order
correlation functions to assess how detailed features of the dynamics
correspond to aspects of the relaxation in glassy systems.  In
particular, Heuer et al. have focused on multiple-time correlation
functions designed to probe {\it relaxation
type}\cite{DoliwaHeuer98,DoliwaHeuer99} as well as {\it rate
memory}\cite{Heuer97} associated with the persistence of slow particle
motion in supercooled liquid systems.  The basic idea of the multiple
time correlation approach is to examine correlations of particle
motion over several time intervals, separating out distance and
directional correlation\cite{HeuerOkun97}. In this fashion, one
effectively devises time filters that extract a particular feature of
the dynamics to be analyzed.  The fundamental building block of the
time correlation functions is the (real part of) tagged particle
density at time interval $\Delta t_{01} = t_1 - t_0$ defined to be
\begin{eqnarray}
f(t_0,t_1) &=& \cos{\left( \vek \cdot (\mathbf{r} (t_1) - \mathbf{r} (t_0))\right)} \equiv 
 \cos{\left( \vek \cdot \Delta \mathbf{r}_{01}\right)}\nonumber \\
&=& \mbox{Re}\left( \hat{N}_{1}(\vek, t_1) \hat{N}_{1}^{*}(\vek, t_0) \right)
\end{eqnarray}
whose ensemble average gives the incoherent scattering function
$F_{s}(\vek, \Delta t_{01})$. Intuitively, $F_{s}(\vek ,\Delta t)$
measures the fraction of particles moving a distance less than $2\pi /
k$ over the time interval $\Delta t$, and hence $f(t_0,t_1)$ can be
viewed as a time-filter selecting out slowly-moving particles.  The
essential idea in identifying the relaxation type is to consider how
the motion of slow-particles is correlated over subsequent time
intervals.  In the purely heterogeneous scenario, one expects that
motion in subsequent time intervals has no direction dependence (no
back-and-forth motion).  On the other hand, for the purely homogeneous
scenario, one rules out a distance dependence in subsequent time
intervals to exclude the presence of different mobilities.  In order
to characterize these limits, it is helpful to define the three-time
correlation function\cite{HeuerOkun97}
\begin{eqnarray}
F_{3}(\Delta t_{01},\Delta t_{12}) &=& \average{f(t_0,t_1) f(t_1,t_2)} \nonumber \\
&=& \average{ \cos\left( \vek \cdot \Delta \mathbf{r}_{01} \right)  \cos\left( \vek \cdot \Delta \mathbf{r}_{12} \right)}.
\eql{F3}
\end{eqnarray}
In the homogeneous limit, the projected distance $\hat{\vek} \cdot \Delta
\mathbf{r}_{12}$ along $\hat{\vek}$ is independent of $\vek \cdot \Delta \mathbf{r}_{01}$, and hence
the three-time correlation function factors to
\begin{eqnarray}
F_{3}(\Delta t_{01},\Delta t_{12}) &=& \average{f(t_0,t_1)}\average{f(t_1,t_2)} \nonumber \\
&=& F_{s}(\vek, \Delta t_{01}) F_{s}(\vek, \Delta t_{12}),
\end{eqnarray}
which suggests defining an indicator function for homogeneous dynamics
\begin{align}
F_{3}^{hom}(\vek, \Delta t_{01}, \Delta t_{12}) =\,&
F_3(\Delta t_{01},\Delta t_{12})
\nonumber \\
&-  F_{s}(\vek, \Delta t_{01}) F_{s}(\vek, \Delta t_{12})
\eql{homoIndicator}
\end{align}
that vanishes in the homogeneous limit.  On the other hand, in the
case of purely heterogeneous dynamics, consider the
indicator\cite{HeuerOkun97}
\begin{align}
F_{3}^{het}(\vek, \Delta t_{01}, \Delta t_{12})=& F_{s}(\vek, \Delta
t_{10} + \Delta t_{12})
\nonumber\\& - 
F_{3}(\Delta t_{01},\Delta t_{12}) 
\eql{heteroIndicator}
\\
=& - \average{ \sin\left( \vek \cdot \Delta \mathbf{r}_{01} \right) \sin\left( \vek \cdot \Delta \mathbf{r}_{01} \right)}.
\nonumber
\end{align}
Since the direction of the motion in subsequent time intervals is not
correlated in the heterogeneous limit, the right hand side of
\eq{heteroIndicator} vanishes.  Note that both indicator
functions make use of the three-time correlation function of the
tagged particle density and can be expressed in terms the
multiple-time propagators of the previous section as
\begin{align}
F_{3}(\vek,&\Delta t_{01}, \Delta t_{12}) = \frac{1}{2}
F_{s}(\vek,\Delta t_{01} + \Delta t_{12}) 
\eql{F3equality}\\
&+ 
\frac{1}{2} \inprod{\hat{N}_{1}(\vek , t_0+t_1+t_2)\hat{N}_{1}(-2\vek , t_0+t_1)}{\hat{N}_{1}(\vek , t_0)},
\nonumber 
\end{align}
where the second term on the right hand side is a special case of the
more general propagator $G_{111}^{(2)}$ in \eq{leading} defined
with $\veq = -2\vek$ and $t_0=0$ and the tagged particle densities
corresponding to the number density.  Both measures of relaxation type
have been successfully tested\cite{Qianetal99} on simple 1-dimensional
model systems in which the dynamical rules governing motion of a
particle are constructed to be inherently heterogeneous (an ensemble
of particles each moving with constant but different jump rates) or
homogeneous (a collection of particles in which particles hop between
sites with two different site-dependent rates).

As the function $f(t_0,t_1)$ acts as a slow-dynamics filter, it can be
used as a means of selecting a sub-ensemble of the full system.  As an
alternative to $F^{hom}_3$ and $F^{het}_3$ in studying the
heterogeneous nature of the dynamics, one can then examine examine how
long particles that are initially in the slow-dynamics ensemble remain
in this ensemble to get an idea of how long solid-like domains persist
in supercooled and glassy systems.  Such filters are also useful to
try to rigorously map deterministic systems onto simplified models of
glassy behavior, such as facilitated spin
models\cite{FredericksonAndersen84,FredricksonAndersen85,Pittsetal00,PittsAndersen01,RitortSollich03,VanZonSchofield05}.
A suitable measure of the lifetime of solid-like domains can be
defined by constructing the four-time correlation function
\begin{eqnarray}
\tilde{C}^{(4)}(t_0,t_1,t_2,t_3) &=& \average{f(t_0,t_1)f(t_2,t_3)} \nonumber \\
&=& \average{\cos\left( \vek \cdot \Delta \mathbf{r}_{01} \right)  \cos\left( \vek \cdot \Delta \mathbf{r}_{23} \right)}.
\nonumber
\end{eqnarray} 
Generally, it is sufficient to look at a time filter over a fixed
period $\Delta t = t_1 - t_0 = t_3 - t_2$, where the waiting interval
$t_2=t_1+\tau$ between subsequent applications of the time filter is
$\tau$.  For large waiting times $\tau$, once expects that only a
random selection of the particles initially in the slow-ensemble
remain in the slow ensemble so that $\tilde{C}^{(4)} \rightarrow
F_{s}(\vek,\Delta t)^{2}$ as $\tau \rightarrow \infty$.  It therefore is
logical to focus on the fluctuation of $f(t_0,t_1)$ defined as
\begin{align}
C^{(4)}(\Delta t, \tau) = &\average{f(0,\Delta t)f(\tau+\Delta t, \tau+2\Delta t)} -F_s(\vek,\Delta t)^2.
\nonumber
\end{align}
If $\Delta t$ is chosen to be shorter than the inverse of the (typical) relaxation rate of
solid-like domains then the time scale $\tau$ at which this decays to
zero can be interpreted as the domain relaxation time.

Note that this quantity is related to the $G_{1111}^{(3)}$, expressed in
\eq{G3leading} and \eq{G3mc}, via
\begin{align}
  C^{(4)}(\Delta t,\tau) = &\frac14\Big[
G^{(3)}_{\kappa'\kappa\kappa\kappa}(\Delta t,\tau,\Delta t)
+
G^{(3)}_{\kappa\kappa'\kappa\kappa}(\Delta t,\tau,\Delta t)
\nonumber\\&+
G^{(3)}_{\kappa\kappa'\kappa'\kappa'}(\Delta t,\tau,\Delta t)
+
G^{(3)}_{\kappa'\kappa\kappa'\kappa'}(\Delta t,\tau,\Delta t)
\Big]
\nonumber\\&
-F_s(\vek,\Delta t)^2,
\eql{domainIndicator}
\end{align}
where $\kappa=\{N_1(\vek)\}$ and $\kappa'=\{N_1(-\vek)\}$.

\subsection{Calculation of domain relaxation rate and relaxation type
  indicators}
Application of the mode-coupling theory of multiple-time correlation
functions developed in \refsec{Formalism} to evaluate the domain
relaxation rate via \eq{domainIndicator} or the relaxation type
indicators defined in \eq{homoIndicator} and
\eq{heteroIndicator} requires complete specification of the slow
basis set variables.  As the indicators are of significant interest in
dense supercooled liquids in which $F_{s}(\vek,t)$ exhibits
non-exponential decay on molecular length scales, the relevant slow
modes must describe the long-time evolution of density fluctuations
for wave vectors $k$ near the peak in the static structure
factor. Clearly the dynamics at such short length scales is outside
the regime of hydrodynamic theory for which one has a good idea of
what constitutes the slow modes of the system.  For dense systems,
however, there is solid evidence from the theory of hard sphere
liquids\cite{DeSchepperCohen80,Kamgar-Parsietal87} of the existence of
short-wavelength ``collective'' modes that are significantly slower
than other ``kinetic'' modes of the system.  These collective modes
are generalizations of the hydrodynamic tagged particle and heat
density modes to finite wave vectors.  The application of the
mode-coupling theory outlined here to molecular length scales is
challenging due to the difficulty in evaluating the contribution of
the fluctuating forces $\varphi_{\alpha} (t)$ to the coupling vertices
$\bar{M}_{\alpha \beta \gamma}$, and requires new input from either
kinetic theory or simulation.  Work along these lines is in progress.

In order to get a feeling of what the mode-coupling predictions of the
correlation functions defined above look like, we focus on a
moderately dense system (in fact relatively dilute compared to a
glass) and examine these functions in the hydrodynamic limit, as was
done in \refcite{VanZonSchofield02b}.  For such a system, it is
sufficient to let the set of slow modes be composed of the tagged
particle number density fluctuations $\hat{N}_{1}(\vek)$ and the
collective hydrodynamic densities, namely the number density
fluctuations $\hat{N}(\vek)$, the longitudinal momentum density
$P_l(\vek)=\hat{\vek}\cdot \mathbf{P}(\vek)$, and the orthogonalized
energy density fluctuations $H(\vek)$ (see
\refcite{VanZonSchofield02b} for the precise definitions of these
variables for a hard sphere system).  For this specific choice of
basis set, the time-derivative of the tagged particle number density
{\it does} have a fluctuating component $\varphi_{N_1}^{s}(\vek,t)$
that contributes to the $\bar{M}^{sss}_{111}$ vertex.  However since
the time derivative of $\hat{N}_{1}(\vek)$ is proportional to
$k=|\vek|$, one expects these ``dissipative'' contributions to be
relatively unimportant in the hydrodynamic limit compared to the
non-dissipative couplings $\average{Q^{s}_{\alpha}Q_{\beta}^{s}
Q_{\hat{\delta}}^{s}}*K_{\hat{\delta}\delta}^{ss-1}$.  The
higher-order vertices necessary to calculate the multiple-time
correlation functions for the indicator functions to leading $N$-order
are therefore
\begin{align}
\bar{M}_{111}^{sss} &= \average{\hat{N}_{1}(\vek - \veq) \hat{N}_{1}(\veq) \hat{N}_{1}^{*}(\vek)} + O(k_0^2) \nonumber \\
&= 1 + O\left( k_0^2 \right) \eql{M111} 
\end{align}
\begin{align}
\bar{M}_{212}^{sss} &= 
\average{Q^{s}_{2}(\vek - \veq - \veq_1,\veq_1) \hat{N}_{1}(\veq) 
Q_{2}^{s*}(\vek - \veq_1'',\veq_1'')}
\nonumber\\&\;\;\;\;
*K_{22}^{ss-1}(\vek - \veq_1'',\veq_1'';\vek - \veq_1',\veq_1') \nonumber \\
& \; \; \; \; \; \; + O\left( k_0^2 \right) \nonumber \\
&= K_{11}^{cc}(\veq_1)*K_{11}^{cc-1}(\veq_1)
\delta_{\veq_1\veq_1'}  
\nonumber\\&
= \delta_{\veq_1\veq_1'} + O \left( k_0^2 \right),
\label{M212}
\end{align}
where we have used the factorization properties
\cite{SchofieldOppenheim92} of multiple-point correlation functions.
In the equations above, $k_0$ represents the largest wave vector
present in the correlation function.  The static part of the
multiple-point vertices $\bar{M}_{211}^{sss}$ and
$\bar{M}_{112}^{sss}$ coming from the time integral of \eq{part1}
vanish since $Q_{1}^{s}(\vek)Q_{1}^{s}(\vek_1)=Q_{1}^{s}(\vek +
\vek_1)$ and $\langle Q_{2}^{s} Q^s_{1} \rangle = 0$ by construction.
The lowest-order contribution in wave vector to these vertices
therefore comes from the time integral of \eq{part2} and is $O(k_0^2)$.

\subsubsection{Relaxation type}
Combining these results with the leading $N$-order expansion terms of
$G_{111}^{(2)}$ and insertion into the expression for $F_{3}$ in
\eq{F3equality} yields
\begin{align}
F_{3}(\vek,t_1,t_2) = \,&\frac{1}{2} F_{s}(\vek,\Delta t_{01} + 
\Delta t_{12}) \nonumber \\
&+ \frac{1}{2} F_{s}(\vek,\Delta t_{12})F_{s}(\vek,\Delta t_{01}) \nonumber \\
&+ \frac{1}{2}F_{3}^{mc}(\vek,\Delta t_{01},\Delta t_{12}) \eql{F3derived}
\end{align}
where $F_{3}^{mc}$ is the first mode-coupling contribution to $F_{3}$.
Note that from \eq{correction}, we see the mode-coupling
corrections involve the evaluation of terms such as
\begin{align}
\sum_{i,\veq}& G_{12}^{ss}(\hat{N}_{1}(\vek); \hat{N}_{1}(\vek - \veq) B_i(\veq); \Delta t_{12})  \nonumber \\
&*G_{21}^{ss}(\hat{N}_{1}(\vek - \veq) B_i(\veq);\hat{N}_{1}(\vek);\Delta t_{01}).
\eql{F3modeCoupling}
\end{align}
In \eq{F3modeCoupling}, the sum extends over the three
hydrodynamic collective variables $B_i(\veq)$ and $G_{12}^{ss}$,
$G_{21}^{ss}$ denote the multiple-point mixed tagged/collective
correlation function
\begin{align}
G_{12}^{ss}(&\hat{N}_{1}(\vek); \hat{N}_{1}(\vek - \veq) B_i(\veq); t) \nonumber \\ &\equiv \left\langle
\hat{N}_{1}(\vek,t)Q_{2}^{s;N_1B_i*}(\vek - \veq, \veq) \right\rangle / \left\langle B_{i}(\veq) B^{*}_{i}(\veq) \right\rangle 
\nonumber
\end{align}
provided the collective densities are orthogonal $\langle B_i(\vek)
B^{*}_{j}(\vek) \rangle = \delta_{ij} \langle B_i(\vek) B^{*}_{i}(\vek)
\rangle$. Similarly, $G_{21}$ is defined as
\begin{align}
G_{21}^{ss}(&\hat{N}_{1}(\vek - \veq) B_i(\veq); \hat{N}_{1}(\vek); t) \nonumber \\ 
&\equiv \left\langle
Q_{2}^{s;N_1B_i}(\vek - \veq, \veq;t) \hat{N}^{*}_{1}(\vek) \right\rangle.
\eql{G21eq1} 
\end{align}
Using the mode-order expansion \eq{physicalexpansion}, the
multiple-point correlation function $G_{21}$ can be approximately
written as
\begin{align}
&G_{21}^{N_1B_i;N_1}(\vek - \veq, \veq;t)  = \int_{0}^{t} d\tau \; F_{s}(\vek- \veq,t-\tau) \eql{G21eq2} \\
& \qquad \times \sum_j G_{11}^{B_i B_j}(\veq,t-\tau) 
M_{21}^{N_1B_j;N_1}(\vek-\veq,\veq;\vek) F_{s}(\vek,\tau) ,\nonumber
\end{align}
with a similar expression for $G_{12}$.  Note that in
\eq{G21eq2}, there is an implicit sum over collective mode
index $j$.  In practice, it is often convenient to work in a basis set
in which the matrix of collective linear-linear (normalized)
correlation functions is diagonal.  In the hydrodynamic regime, this
corresponds to choosing the $B_i$ set to be composed of the
hydrodynamic sound and heat modes.

Returning to the problem of calculating relaxation type, inserting
\eq{F3derived} into the definition of the indicator functions
yields
\begin{align}
F_{3}^{hom}(\vek,\Delta t_{01}, \Delta t_{12}) &= \frac12 [\Delta
F_{2}(\Delta t_{01}, \Delta t_{12}) + F_{3}^{mc} ]
\eql{Hom2}\\
F_{3}^{het}(\vek,\Delta t_{01}, \Delta t_{12}) &=\frac12 [ \Delta F_{2}(\Delta t_{01}, \Delta t_{12}) -  F_{3}^{mc}]
\eql{Het2}
\end{align}
where
\begin{align}
\Delta F_{2}(\Delta t_{01}, \Delta t_{12}) =  \,&
F_{s}(\vek,\Delta t_{01}+\Delta t_{12}) 
\eql{Delta F2} \\
& - F_{s}(\vek,\Delta t_{01})
F_{s}(\vek,\Delta t_{12}) \nonumber .
\end{align}
In the hydrodynamic limit and ignoring mode-coupling effects, one
expects that an exponential decay of the intermediate scattering
function of the form $F_{s}(\vek,t) \sim \exp (-D|\vek|^2t)$, where $D$ is the
self-diffusion coefficient.  In this case, $\Delta F_{2} = 0$ and both
indicators are approximately zero.  On the other hand, if $F_{s}(\vek,t)$
has a non-trivial relaxation profile, perhaps of the form of a
stretched exponential $F_{s}(\vek,t) \sim \exp\{ -[t/\tau
(\vek)]^{\tilde{\beta}}\}$ where $\tilde{\beta}$ is the stretching
exponent, then $\Delta F_{2} \neq 0$.  It is therefore evident that
mode-coupling effects are absolutely essential to distinguish between
homogeneous and heterogeneous types of non-exponential relaxation.
Although this result is obvious since mode-coupling must be invoked to
give rise to non-exponential relaxation in the first place, it is the
{\it difference} in the time dependence of $\Delta F_{2}$ and the
mode-coupling corrections $F_{3}^{mc}$ that allows one to distinguish
between the two relaxation types.  On the molecular scale, one also
anticipates a contribution to these expressions from the dissipative
part of $\bar{M}_{111}$ that corrects the simple factorization result
at lowest $N$-order
\begin{align}
& \left\langle \hat{N}_{1}(\vek - \veq, t_1 + t_2) \hat{N}_{1}(\veq,t_1) \hat{N}_{1}^{*}(\vek) \right\rangle \nonumber \\
& \; \; \; \; \; \; \; \approx
F_{s}(\vek - \veq, t_2) F_{s}(\vek , t_1)
 \left[ 1 + O \left( k_{0}^{2} \right) \right].
\eql{factorization}
\end{align}
Note that the effect of these corrections is not to modify the time
behavior but to modify the wave vector dependence on the right-hand
side of \eq{factorization}.  This modification only makes sense
when $t_1$ and $t_2$ are large compared to the microscopic relaxation
time $t_m$ corresponding to the time scale after which the
instantaneous approximation implicit in \eq{instantmultiQ} is valid.

\subsubsection{Domain relaxation}

The rate of domain relaxation can be similarly calculated using the
$N$-ordering scheme to simplify the multiple-time correlation
functions in \eq{domainIndicator}.  Using the form of the higher-order
vertices in the hydrodynamic limit and $G_{00}=1$, $G^{ss}_{11}(\hat
N_1(\vek=0);\hat N_1(\vek=0);\tau)=0$, and $G^{ss}_{11}(\hat
N_1(\vek);\hat N_1(\vek);\Delta t)=F_s(\vek,\Delta t)$, one obtains
\begin{align}
G^{(3)}_{\hat\kappa\kappa\kappa\kappa}(\Delta t,\tau,\Delta t)
\approx F_s(\vek,\Delta t) F_s(\vek,\Delta t).
\end{align}
The same holds for the other four time correlations in
\eq{domainIndicator}, therefore the leading orders of the
multiple-time correlations are canceled by the last term in
\eq{domainIndicator}, so that to leading order $C^{(4)}(\Delta
t,\tau)$ is zero.  The first non-zero contribution arises at first
mode-coupling order, as given by \eq{G3mc}. However, the terms in that
equation that involve $G_{21}(\hat N_1(\veq),B_i(-\veq);\hat
N_1(0);t)$ are zero because $\hat N_1(0)=0$.  Thus, only the last term
from \eq{G3mc} survives. Taking together the contributions from the
four higher-order terms in \eq{domainIndicator} and using that
$G^{ss}_{22}$ factorizes to leading order, gives the mode coupling
result
\begin{align}
C^{(4)}(\Delta t,\tau) = \mbox{Re}&
\bigg(\sum_{i,j,\veq}
G_{12}^{ss}(\hat N_1(\vek);\hat N_1(\vek-\veq)B_i(\veq);\Delta t)
\nonumber\\&\times
F_s(\veq,\tau) G_{11}^{B_iB_j}(\veq,\tau) 
\nonumber\\&\times
G_{21}^{ss}(\hat N_1(\vek-\veq)B_j(\veq);N_1(\vek);\Delta t)
\bigg)
\nonumber\\&+O(k_0^2)
,
\end{align}
where \eq{G21eq2} may be used for $G_{12}^{ss}$ and $G_{21}^{ss}$.

As this result already somewhat suggests, for dense, supercooled
liquids, one anticipates that the conversion time of solid-like
domains to liquid-like domains of typical length scale $l \sim 2\pi/k$
also occurs on time scales corresponding to the wave vector-dependent
$\alpha$-relaxation time of $F_{s}(\vek,t)$.

\subsection{Numerical analysis of multiple-time correlation functions of tagged densities in a hard sphere system}
In order to validate the mode-coupling theory approach to
multiple-time correlation functions of collective densities, extensive
simulations of a hard sphere system at moderate densities were carried
in \refcite{VanZonSchofield02b}.  It is straightforward to extend
these simulations to incorporate calculations of tagged particle
densities to test the simple factorization result for multiple-time
correlation functions of the tagged particle number density in
\eq{factorization} and to examine how well the theory predicts
higher-point correlation functions of mixed tagged/collective
densities that are necessary to compute the mode-coupling corrections
to the leading $N$-order factorization.  All simulation results
presented in this section were obtained using the simulation
methodology described in detail in \refcite{VanZonSchofield02b} on a
hard sphere system of a relatively low reduced density
$\rho^{*}=\rho/\rho_c=0.1$, where $\rho_c$ is the density at close
packing.  The size of the periodic system was chosen to have cubic box
lengths $L_x=L_y=L_y= 47.3361$ such that the simulation system
contained $N=15000$ hard-sphere particles of mass $m=1$ and diameter
$a=1$ at the chosen density.  For this system size, the smallest
dimensionless wave vector $k_0a = 2\pi a/L_x=0.132736$ so that all
quantities examined are roughly in the hydrodynamic regime.  For this
system, the mean collision time calculated from Enskog
theory\cite{ChapmanCowling} is approximately $t_e=1.42417$ at an
inverse temperature $\beta = 3$, while the mean-free path is
$l_e=1.85561$ so that $k_0 l_e =0.246305$.  Under these conditions,
the estimated relaxation time of $F_{s}(k_0,t)$ is $\tau(k_0) \sim
(Dk_0^2)^{-1} \approx 55 t_e$, where $D= 0.725586$ is the
self-diffusion coefficient calculated from the Enskog theory result
\begin{equation}
D = \frac{3w_s}{8 \sqrt{\beta m \pi}g(a)a^2},
\end{equation}
where $w_s=1.01896$ and $g(a)$ is the radial distribution function at
contact.  For a hard-sphere system, $g(a)$ can be estimated using the
Carnahan-Starling equation of state\cite{CarnahanStarling69} and the
expression for the pressure $p$ of a hard-sphere system
\begin{equation}
\frac{\beta p}{\rho} = \frac{1+\bar{\eta}+\bar{\eta}^{2}-\bar{\eta}^3}{(1-\bar{\eta})^{3}} = 1+ b \rho g(a),
\end{equation}
where $b=2\pi a^3/3$ and $\bar{\eta} = \pi \rho a^3/6$.  The
simulations were run for at total time of approximately $1200$
$\tau(k_0)$ to insure reasonably good statistics.  Following
\refcite{VanZonSchofield02b}, statistical uncertainties were
estimated using the symmetry properties of the correlation functions.
In this approach, the statistical uncertainty for a real correlation
function was constructed from a histogram of the values of the
imaginary part of the complex correlation function, which vanishes on
average, to determine the $96 \%$ confidence intervals.  To simplify
the comparison between theoretical predictions and the simulation
results, all wave vectors were taken to be co-linear so that $\vek
\cdot \veq = kq$.  To further improve statistics, the wave vectors
were independently taken along the three principal directions
$\hat{x}$, $\hat{y}$ and $\hat{z}$ of the cubic simulation box and
averaged.

\subsubsection{Two time-interval correlation function}
\begin{figure}[b]
\vspace{.2in}
\centerline{\includegraphics[width=0.47\textwidth]{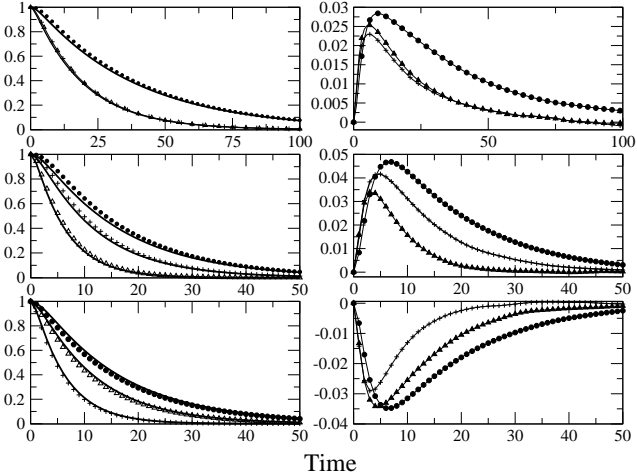}}
\caption{A comparison of the multiple-time correlation function
$g^{(2)}$, the factored form $g^{(2)}_{f}$ and their difference as a
function of wave vector pair $(m,n)$ and time cross-section
$(t_1,t_2)$.  The left hand side panels contain the simulation results
for $g^{(2)}$ (unconnected symbols) and $g^{(2)}_{f}$ (lines), and the
right hand side panels contain plots of $g^{(2)}-g^{(2)}_f$.  In all
panels, the unconnected dots, crosses and triangles correspond to the
simulation results for time cross sections $(t,t)$, $(3t,t)$ and
$(t,3t)$, respectively.  The results in the top, middle and bottom
rows are for wave vector sets $(1,2)$, $(1,3)$ and $(2,1)$,
respectively.  For clarity in the figures, the $96\%$ confidence
intervals, estimated to be roughly $0.002$, have been omitted.}
\label{multitime}
\end{figure}

One of the central results of \refsec{Applications}, the simple
factorization of multiple-time correlation functions of the tagged
particle number density (see \eq{factorization}) is simple to
verify numerically.  To validate this prediction, numerical
calculations of the correlation function $g^{(2)}(t_1,t_2)=\langle
\hat{N}_{1}(\vek - \veq, t_1 + t_2) \hat{N}_{1}(\veq,t_1)
\hat{N}_{1}^{*}(\vek) \rangle$ were computed as function of wave
vector combinations $(\vek,\veq)=(m,n)\vek_0$ for $m$ and $n$ values
ranging from $1$ to $3$ with $m \neq n$.  To simplify the comparisons,
three different time cross-sections $(t_1,t_2)$ for each pair $(m,n)$
were calculated, namely $(t,t)$, $(t,3t)$ and $(3t,t)$.  These
particular choices of $t_1$ and $t_2$ are the simplest to implement
when simulation data is stored in standard linear array data
structures.  The results, shown in \fig{multitime}, show that the
product $g^{(2)}_{f}=F_{s}(\vek-\veq,t_2)F_{s}(\vek,t_1)$ approximates
$g^{(2)}(t_1,t_2)$ very well for all choices of wave vector
combinations and all time cross-sections.  However, given that the
statistical uncertainties for both $g^{(2)}$ and $g^{(2)}_{f}$ are on
the order of $0.002$, it is clear that the numerical results indicate
a small but systematic difference between these two quantities which
reaches a maximum a short times $t \sim 5 t_e$, as can be seen in the
panels on the right hand side of \fig{multitime}.

This difference should be well-reproduced by the mode-coupling
correction term in \eq{F3modeCoupling}, although this has not
yet been verified.

\subsubsection{Multiple-point correlation function}
Although it is certainly possible to calculate the leading order
mode-coupling correction term in \eq{F3modeCoupling}
analytically using hydrodynamic forms for the collective correlation
functions $G_{11}^{cc}(t)$, the calculation is tedious due to the sum
over different collective modes $B_i$.  However in order to validate
the applicability of the mode-coupling theory in the calculation of
the first order correction terms $F^{mc}_{3}$, we examine the
quality of the predicted form of the multiple-point correlation
function
\begin{align}
G_{21}^{ss}(&\hat{N}_{1}(\vek - \veq) \hat{N}(\veq);
\hat{N}_{1}(\vek); t) 
\eql{GNN_N1} \\ 
&\equiv \left\langle
\hat{N}_1(\vek - \veq,t) \hat{N}(\veq,t) \hat{N}^{*}_{1}(\vek)
\right\rangle 
- \frac{S(\veq)}{\langle N \rangle}F_{s}(\vek,t),
\nonumber 
\end{align}
where $S(\veq)=\langle \hat{N}(\veq) \hat{N}^{*}(\veq) \rangle$ is the
static structure factor, against direct numerical calculation (note:
in \eq{GNN_N1} we took the thermodynamic limit).  From
\eq{G21eq2}, we see that this correlation function is
approximately given by
\begin{align}
&G_{21}^{N_1N;N_1}(\vek - \veq, \veq;t)  = \int_{0}^{t} d\tau \;
  F_{s}(\vek- \veq,t-\tau) 
\eql{G21eq3} \\
& \qquad \times G_{11}^{NB_j}(\veq,t-\tau) 
M_{21}^{N_1B_j;N_1}(\vek-\veq,\veq;\vek) F_{s}(\vek,\tau) ,\nonumber
\end{align}
where $B_j$ runs over the set $\{N,P_l,H\}$ or some variant of it.
Note that with this choice of collective densities, the computation of
\eq{G21eq3} requires the calculation of $3$ coupling vertices
$M_{21}^{N_1N;N_1}$, $M_{21}^{N_1P_l;N_1}$, and $M_{21}^{N_1H;N_1}$,
as well as the input of the collective linear-linear time correlation
functions $G_{11}^{NN}$, $G_{11}^{NP_l}$, and $G_{11}^{NH}$.  In
principle, analytical forms of the $G_{11}$ time correlation functions
can be utilized in the hydrodynamic regime with transport coefficients
either fitted from simulation data or taken from kinetic theory.
Given the simplicity and ease of numerically calculating the $G_{11}$
with excellent precision, we have chosen to evaluate the convolution
integrals in \eq{G21eq3} by numerically integrating simulation
data using a version of Simpson's rule that allows interpolation of
data.  From time-inversion symmetry, $M_{21}^E = 0$ for $B_j=N$ or
$H$, and it is evident that the only coupling at ``Euler'' order
arises for $B_j=P_l$.  However, as noted for the case of
multiple-point correlation functions of purely collective densities in
\refcite{VanZonSchofield02b}, the inclusion of the
additional couplings to $N$ and $H$ arising at ``dissipative'' order
$M_{21}^D$ is essential if quantitatively accurate predictions are
desired.  It may appear at first glance that the dissipative
contributions are negligible in the limit of small wave vectors since
they introduce additional factors of $k_0$.  In fact the overall order
of the multiple-point correlation functions is determined by a wave
vector-dependent prefactor multiplied by the convolution of the
two-point, single time interval correlation functions $G_{11}$.  The
time convolution of these correlation functions can give rise to
additional factors of wave vector depending on their symmetry
properties, so that the overall contribution of the $P_l$ and $N$ or
$H$ coupling are comparable in magnitude.

\begin{figure}[b]
\vspace{.2in}
\centerline{\includegraphics[width=0.47\textwidth]{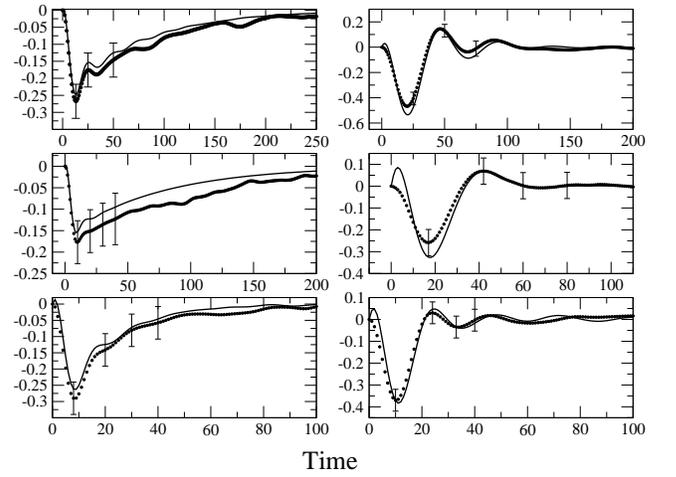}}
\caption{A comparison of simulated (unconnected dots) and predicted
(solid lines) values of $G_{21}(t)$ as a function of $t$.  The results
in the top row correspond to wave vectors pairs $(1,2)$ and $(2,1)$
from left to right, the middle row correspond to $(1,3)$ and $(3,1)$,
and the bottom row to $(2,3)$ and $(3,2)$.}
\label{G21fig}
\end{figure}

The evaluation of the dissipative contribution to coupling vertices
$M_{21}^{N_1N;N_1}$ and $M_{21}^{N_1H;N_1}$ requires external input.
In the appendix, these vertices are evaluated in the small wave vector
limit by relating the dissipative vertices to the Enskog
self-diffusion coefficient and its derivatives with respect to
thermodynamic parameters.  The predictions are therefore free of any
adjustable parameters and constitute a rigorous test of the
mode-coupling theory.  The results of the comparison are shown in
\fig{G21fig}.

Note that although the data are a bit noisy, the theoretical
predictions generally fall within the confidence intervals of the
simulated data except at very short times ($t \sim 2 t_e$) where one
expects the theory to break down due to the instantaneous
approximation of the coupling vertices.

\section{Summary}
\label{Summary}

In this paper, a mode-coupling theory was presented in which multiple
point and multiple-time correlation functions for collective densities
and tagged particle densities are expressed in terms of ordinary
two-point, single-time interval correlation functions and a set of
vertices. The theory developed here does not assume that fluctuating
forces (noise) are Gaussian distributed and, in principle, does not
require an ansatz to obtain self-consistent equations.  Furthermore,
unlike kinetic theories, it is not restricted to low densities and
should be applicable to dense fluids where cooperative motions of
particles and collective modes are important.

The formalism is based on projection operator techniques, which, for
ordinary two-point, single-time correlation functions, lead to a
generalized Langevin equation in which the memory function decays on a
microscopic time scale (\eq{memorydiff}).  The simple extension of the
projection operator formalism to multiple-time correlation functions
of tagged particle densities is complicated by the fact that the
fluctuating forces appearing in the generalized Langevin equation do
not obey Gaussian statistics.  Furthermore, multiple-time correlations
of the fluctuating force can in fact have a slow decay when the time
arguments of these forces become comparable.  In order to treat
multiple-time correlation functions of fluctuating forces properly,
the correlation functions were manipulated so that the time arguments
of all fluctuating forces appearing in the correlations were
guaranteed to be well-separated, ensuring that all memory functions
which arise in the mode-coupling theory decay to zero on a molecular
time scale.  This construction allows equations which are local in
time to be obtained which relate the multiple-time correlation
function to two-time but multiple-point correlations coupled by
essentially time-independent vertices.  These expressions, in turn,
can be written as convolutions of two-point, single time interval
``propagators'' coupled by time-independent vertices.  These
propagators can either be taken directly from experiment, simulation,
or can be solved self-consistently within the mode-coupling formalism.
The vertices, which are composed of a static part (Euler term) and a
generalized transport coefficient, can similarly be calculated from
kinetic theory or taken from molecular dynamics and Monte-Carlo
simulations.

The equations for higher-order correlation functions contain an
infinite sum of terms which can be made tractable for systems with a
finite correlation length by applying a cumulant expansion technique
pioneered by Oppenheim and
co-workers\cite{MachtaOppenheim82,Schofieldetal92,SchofieldOppenheim92}
termed the $N$-ordering method.  The method was applied to obtain the
leading order and first mode-coupling corrections of expressions for
tagged-particle density multiple-time and mixed tagged/collective
particle density multiple-point correlation functions designed to
probe detailed aspects of the relaxation profile of glassy systems.

The mode-coupling theory outlined here is the first step towards a
fully microscopic theory applicable to molecular length scales that
enables one to analyze the mapping of the dynamics in deterministic
glassy systems to stochastic spin model of glasses, as well as to
define sub-ensembles via filters based on dynamical and spatial
properties to probe dynamic heterogeneities, clustering properties and
many other aspects of the dynamics of frustrated systems.

\acknowledgments
It is a pleasure to dedicate this paper to Irwin Oppenheim, whose friendship
we greatly treasure.  This work was supported by a grant
from the Natural Sciences and Engineering Research Council of Canada.
J.S.  would like to thank Walter Kob and the Laboratoire des Collo\"\i
des, Verres et Nanomat\`eriaux in Montpellier, France for their
hospitality and the CNRS for additional funding during the final
stages of this work.

\appendix*

\section{Evaluation of the mode-coupling vertices}

As mentioned in the main text, the vertices required for the
calculation of the mode-coupling correction to the factorization of
the multiple-time correlation function $F_{3}(t_1,t_2)$ are
$M_{21}^{N_1P_l;N_1}$, $M_{21}^{N_1N;N_1}$ and $M_{21}^{N_1H;N_1}$,
where the general form of the vertex $M_{21}^{ss}$ is
\begin{align}
M_{21}^{ss} &= M_{21}^{Ess}+M_{21}^{Dss} \nonumber \\
M_{21}^{Ess} &= \left\langle \calL Q_{2}^{s} Q_{1}^{s*} \right\rangle \cdot K_{11}^{ss-1} \\
M_{21}^{Dss} &= -\int_{0}^{\infty} d \tau \left\langle \varphi_{2}^{s}(\tau) \varphi_{1}^{s*} \right\rangle \cdot K_{11}^{ss-1},
\end{align}
where $M_{21}^E$ is referred to as the ``Euler'' contribution to the
vertex, $M_{21}^{D}$ is the ``dissipative'' part of the vertex, and
$\varphi(t)$ is the fluctuating force defined in
\eq{fluctuatingForce}.

Examining first the Euler contributions $M_{21}^{Ess}$, we note that
since $\left\langle \calL Q_{2}^{s} Q_{1}^{s*} \right\rangle = -
\left\langle Q_{2}^{s} \calL Q_{1}^{s*} \right\rangle$, we have
\begin{align}
M_{21}^{Ess}( &\hat{N}_{1}(\vek-\veq)B(\veq);\hat{N}_{1}(\vek)) = 
\nonumber \\
&\frac{i}{m} \left\langle Q_{2}^{s}(\hat{N}_{1}(\vek-\veq),B(\veq))
\mathbf P_1^*(\vek) \right\rangle \cdot \vek.
\end{align}
>From the time-reversal symmetry properties of the correlation
functions, we see that $M_{21}^{Ess}=0$ for $B=N$ and $B=H$, and for
$B=P_l$ we find
\begin{align}
M_{21}^{Ess}( &\hat{N}_{1}(\vek-\veq)P_l(\veq);\hat{N}_{1}(\vek)) = \frac{i\vek \cdot \hat{\veq}}{\beta}.
\end{align}

Evaluation of the dissipative vertices $M_{21}^{Dss}$ requires the
calculation of the linear $\varphi_1$ and bi-linear $\varphi_{2}$
fluctuating forces.  For the linear fluctuating force, we have (in the
thermodynamic limit)
\begin{align}
\varphi_{1}(\vek, t) &= e^{\calP^\perp\calL t} \calP^\perp \calL \hat{N}_{1}(\vek) \nonumber \\
&=  e^{\calP^\perp\calL t}i \mathbf{P}_1(\vek)\cdot \hat{\vek} /m, 
\end{align}
and the bi-linear fluctuating force is 
\begin{align}
\varphi_{2}(\vek-\veq,\veq; t) &= e^{\calP^\perp\calL t} \calP^\perp \calL Q_{2}^{s}(\vek-\veq,\veq) \nonumber,
\end{align}
where $\calP^\perp$ is a projection operator that projects onto a
subspace orthogonal to that spanned by the multi-linear basis set
$Q_{\alpha}$.  Noting that (in the thermodynamic limit)
\begin{align}
Q_{2}^{s}(\hat{N}_{1}(\vek-\veq)\hat{N}(\veq)) &= \hat{N}_{1}(\vek-\veq)\hat{N}(\veq) -
\frac{S(\veq)}{\langle N \rangle} \hat{N}_{1}(\vek) \nonumber \\
Q_{2}^{s}(\hat{N}_{1}(\vek-\veq)P_l(\veq)) &= \hat{N}_{1}(\vek-\veq)P_l(\veq) \nonumber \\
Q_{2}^{s}(\hat{N}_{1}(\vek-\veq)H(\veq)) &= \hat{N}_{1}(\vek-\veq)H(\veq) 
\end{align}
and $\calP^\perp \hat{N}_{1}(\vek - \veq)\mathbf{P}(\veq) = 0$, we see
that at $t=0$, the bilinear fluctuating forces are
\begin{align}
\varphi_{2}^{s}(\hat{N}_{1}(\vek-\veq)\hat{N}(\veq)) &= \frac{i(\vek-\veq)}{m} \cdot \mathbf{P}_1 (\vek-\veq)\hat{N}(\veq) \nonumber \\
& \qquad \qquad - \frac{S(\veq)}{\langle N \rangle} \frac{i\vek}{m} \cdot \mathbf{P}_{1}(\vek) \nonumber \\
\varphi_{2}^{s}(\hat{N}_{1}(\vek-\veq)P_l(\veq)) &= \frac{i(\vek-\veq)}{m} \cdot \calP^\perp 
\mathbf{P}_1 (\vek-\veq) \mathbf{P}(\veq) \cdot \hat{\veq}
\nonumber \\
&\qquad + \calP^\perp \hat{N}_{1}(\vek - \veq)\calL \mathbf{P}(\veq) \cdot \hat{\veq} \nonumber \\
\varphi_{2}^{s}(\hat{N}_{1}(\vek-\veq)H(\veq)) &= \frac{i(\vek-\veq)}{m} \cdot \calP^\perp 
\mathbf{P}_1 (\vek-\veq) H(\veq)
\nonumber \\
&\qquad + \calP^\perp \hat{N}_{1}(\vek - \veq)\calL H(\veq) \nonumber .
\end{align}
Inserting these results into the expressions for the dissipative
vertices and expanding the resulting expressions in powers of the base
wave vector $k_0$, one finds
\begin{align}
M_{21}^{ssN_1N;N_1} &= - \frac{(\vek - \veq)\cdot \vek}{m^2} \int_0^{\infty} d\tau \left\langle \mathbf{P}^{x}_{1}(\tau )
\mathbf{P}^{x}_{1} \hat{N} \right\rangle \nonumber \\
&\qquad + \frac{k^2S(\veq)}{m^2 \langle N \rangle} \int_0^{\infty} d\tau \left\langle \mathbf{P}^{x}_{1}(\tau) 
\mathbf{P}^{x}_{1} \right\rangle 
+O(k_0^3) \nonumber \\
M_{21}^{ssN_1P_l;N_1} &= O(k_0^3) \nonumber \\
M_{21}^{ssN_1H;N_1} &=- \frac{(\vek - \veq)\cdot \vek}{m^2} \int_0^{\infty} d\tau \left\langle \mathbf{P}^{x}_{1}(\tau )
\mathbf{P}^{x}_{1} H \right\rangle \nonumber \\
&\qquad - \frac{\vek \cdot \veq}{m} \int_0^{\infty} d\tau \left\langle \mathbf{j}_{H}(\tau) \mathbf{P}_{1} \right\rangle 
+O(k_0^3) \nonumber,
\end{align}
where we have defined $i\veq \cdot \mathbf{j}_{H}(\veq) = -2
\beta/\sqrt{6} \, \calP^\perp \calL \hat{E}(\veq)$.  Noting that the
Green-Kubo expression for the self-diffusion coefficient $m^2D =
\int_{0}^\infty dt \, \langle \mathbf{P}^{x}_{1}(t) \mathbf{P}_{1}^{x}
\rangle$ and that $\langle A \hat{N} \rangle = \partial \langle A
\rangle / \partial(\beta \mu)$ and $\langle A \hat{E} \rangle =
-\partial \langle A \rangle / \partial\beta$, where $\mu$ is the
chemical potential of the system, allows us to write
\begin{align}
M_{21}^{ssN_1N;N_1} &= - (\vek - \veq)\cdot \vek \left( \frac{\partial D}{\partial \beta \mu} \right)_{\beta}
+ \frac{S(0)}{\langle N \rangle} Dk^2
+O(k_0^3) \nonumber \\
M_{21}^{ssN_1H;N_1} &=- \frac{(\vek - \veq)\cdot \vek}{\sqrt{6}} \left[ 3  
\left( \frac{\partial D}{\partial \beta \mu} \right)_{\beta} + 2 \beta  
\left( \frac{\partial D}{\partial \beta} \right)_{\beta \mu} \right] \nonumber \\
&\qquad - \frac{\vek \cdot \veq}{m} \int_0^{\infty} d\tau \left\langle \mathbf{j}_{H}(\tau) \mathbf{P}_{1} \right\rangle 
+O(k_0^3) \nonumber.
\end{align}
The dissipative vertex $M_{21}^{ssN_1H;N_1}$ contains a new transport
coefficient $\int_0^{\infty} d\tau \left\langle \mathbf{j}_{H}(\tau)
\mathbf{P}_{1} \right\rangle$ which could, in principle, be evaluated by
doing a short-time expansion or using uncorrelated collisions in a
kinetic theory.  Nonetheless, we expect this contribution to be very
small since $\langle \mathbf{j}_{H}(\tau) \mathbf{P} \rangle$ is strictly
zero due to the projection operator $\calP^\perp$.  In the numerical
work in the main text, we have set this term to zero.

Using the Enskog form for the self-diffusion coefficient and the
Carnahan-Starling equation of state, we find that
\begin{align}
\left( \frac{\partial D}{\partial \beta \mu} \right)_{\beta} &= - \frac{D}{g(a)} \frac{\bar{\eta}(2\bar\eta - 5)}{2(1-\bar\eta)^4}
\frac{S(0)}{\langle N \rangle} \nonumber \\
\left( \frac{\partial D}{\partial \beta} \right)_{\beta \mu} &= -\frac{D}{2\beta} + \frac{D}{g(a)} \frac{\bar{\eta}(2\bar\eta - 5)}{2(1-\bar\eta)^4}
\frac{3S(0)}{2\beta\langle N \rangle} .
\end{align}
Thus, the $M_{21}$ vertices used in the main text are:
\begin{align}
M_{21}^{ssN_1N;N_1} &= \frac{D S(0)}{\langle N \rangle} \left[
\frac{\vek \cdot (\vek - \veq)}{g(a)} \frac{\bar{\eta}(2\bar\eta - 5)}{2(1-\bar\eta)^4}
+ k^2 \right] \nonumber \\
M_{21}^{ss}( \hat{N}_{1}P_l;N_1) &= \frac{i\vek \cdot \hat{\veq}}{\beta} \nonumber \\
M_{21}^{ssN_1H;N_1} &= \frac{D (\vek - \veq)\cdot
  \vek}{\sqrt{6}}. \nonumber 
\end{align}

\end{document}